\documentclass[aps,prl,floatfix,superscriptaddress,twocolumn,footinbib]{revtex4-2}

\usepackage{amssymb}
\usepackage{amsmath}
\usepackage{amsfonts}
\usepackage{appendix}
\usepackage{bm}
\usepackage{braket}
\usepackage{graphicx}
\usepackage{epsfig}
\usepackage{epstopdf}
\usepackage{balance}
\usepackage[dvipsnames]{xcolor}
\usepackage{calc}
\usepackage{natbib}
\usepackage[colorlinks,
            linkcolor=blue,
            anchorcolor=blue,
            citecolor=blue,
            urlcolor=blue]{hyperref}
\usepackage{lipsum}
\usepackage{enumerate}

\begin{document}
\title{Energy Bands of Incommensurate Systems}
\author{Xin-Yu Guo}
	\affiliation{School of Physics and Wuhan National High Magnetic Field Center, Huazhong University of Science and Technology, Wuhan, Hubei 430074, China}
 \author{Jin-Rong Chen}
	\affiliation{School of Physics and Wuhan National High Magnetic Field Center, Huazhong University of Science and Technology, Wuhan, Hubei 430074, China}
  \author{Chen Zhao}
	\affiliation{School of Physics and Wuhan National High Magnetic Field Center, Huazhong University of Science and Technology, Wuhan, Hubei 430074, China}
 
  \author{Miao Liang}
	\affiliation{School of Physics and Wuhan National High Magnetic Field Center, Huazhong University of Science and Technology, Wuhan, Hubei 430074, China}
  \author{Ying-Hai Wu}
  \affiliation{School of Physics and Wuhan National High Magnetic Field Center, Huazhong University of Science and Technology, Wuhan, Hubei 430074, China}
      \author{Jin-Hua Gao}
 \email{jinhua@hust.edu.cn}
	\affiliation{School of Physics and Wuhan National High Magnetic Field Center, Huazhong University of Science and Technology, Wuhan, Hubei 430074, China}
 \author{X. C. Xie}
 \affiliation{Interdisciplinary Center for Theoretical Physics and Information Sciences (ICTPIS), Fudan University
, Shanghai 200433, China}
\affiliation{International Center for Quantum Materials, School of Physics, Peking University, Beijing 100871, China}
 \affiliation{Hefei National Laboratory, Hefei 230088, China}
\begin{abstract}

Energy band theory is a fundamental cornerstone of condensed matter physics. According to conventional wisdom, discrete translational symmetry is mandatory for defining energy bands. Here, we illustrate that, in fact, the concept of energy band can be generalized to incommensurate systems lacking such symmetry, thus transcending the traditional paradigm of energy band.   The validity of our theory is verified by extensive numerical calculations in the celebrated Aubry-Andr\'e-Harper model and a two-dimensional incommensurate model of graphene. Building upon the proposed concept of incommensurate energy bands, we further develop a theory of  angle-resolved photoemission spectroscopy (ARPES) for incommensurate systems, providing a clear physical picture for the incommensurate  ARPES spectra. Our work establishes a comprehensive  energy band theory for incommensurate systems.

\end{abstract}
\maketitle
\emph{Introduction.}---Energy band is one of the central concepts of solid state physics, serving as the standard language for describing the electronic structure of periodic systems. Theoretically, the energy bands are rigorously defined based on Bloch's theorem~\cite{solidstatebook}, which stems from the inherent translational symmetry of the crystal lattice.

Unlike periodic systems, the incommensurate (or quasiperiodic) systems refer to electrons in a potential with two or more competing incommensurate periods, leading to the absence of overall translational symmetry. The quasiperiodic potential can induce many unique features, attracting significant research interest over the past few decades. For instance, the Aubry-André-Harper (AAH) model~\cite{PG.Harper_1955,aubry1980analyticity}, one dimensional atomic chain subjected to a quasiperiodic potential, may be the first celebrated example of the incommensurate systems. This model, extensively studied to explore the localization phenomena~\cite{PhysRevB.28.4272,PhysRevLett.61.2144_1988,PhysRevLett.61.2141,PhysRevB.41.5544_1990,PhysRevLett.98.130404,roati2008anderson,PhysRevLett.103.013901_2009,modugno2009exponential,PhysRevLett.109.106402_2012,PhysRevLett.108.220401_2012,schreiber2015observation,PhysRevB.96.085119,PhysRevLett.120.160404,PhysRevLett.125.196604_2020} and Hofstadter spectrum problems~\cite{PhysRevB.14.2239_1976,hatsugai1990energy,satija2016butterfly}, has also garnered significant attention in the field of mathematics~\cite{hakan1997, avila2005,avila2015,zhouqi2015}.  Meanwhile, the recently intensively studied moir\'{e} systems~\cite{cao2018correlated,cao2018unconventional,Macdonald20112,koshino2018,wu2018,PhysRevLett.123.036401_2019,PhysRevB.85.195458_2012,PhysRevLett.122.257002_2019,PhysRevLett.109.126801,Guo2018,PhysRevB.102.035136_2020,PhysRevLett.125.257602_2020,helin121406_2020,PhysRevLett.127.197701_2021,liu2020tunable,shen2020correlated,lee2019theory,PhysRevB.99.235406,chen2021electrically,polshyn2020electrical,PhysRevLett.125.116404,ma2021topological,PhysRevLett.127.166802,zeng_thermodynamic_2023,cai2023signatures,park2022robust,PhysRevLett.128.026403,zhang2022promotion,PhysRevB.105.195422,PhysRevB.108.195119}, e.g.~twisted bilayer graphene (TBG) ~\cite{Macdonald20112,koshino2018,cao2018correlated,cao2018unconventional,wu2018,PhysRevLett.123.036401_2019,PhysRevB.85.195458_2012,PhysRevLett.122.257002_2019,PhysRevLett.109.126801,Guo2018,PhysRevB.102.035136_2020,PhysRevLett.125.257602_2020,helin121406_2020,PhysRevLett.127.197701_2021}, are typical two dimensional (2D) incommensurate systems,  exhibiting exotic moir\'{e} flat band physics~\cite{andrei2021marvels,mak2022semiconductor,du2023moire}.  

The energy band concept, as well as Bloch's theorem, is in principle inapplicable to incommensurate systems, due to the lack of translational symmetry. This poses a significant theoretical challenge, especially for those truly incommensurate structures that exhibit marked deviation from commensurate cases, e.g.~moir\'{e} quasicrystals~\cite{lubin2012high,PhysRevB.93.201404_2016,ahn2018dirac,zhoushuyun2018,yu2019dodecagonal,PhysRevB.102.045113,Ye2020,PhysRevResearch.4.013028_2022,uri2023superconductivity,li2024tuning}, since energy band theory becomes invalid at all.  However, in experiments, dispersing band states can still be clearly observed in various incommensurate systems,
through ARPES measurements ~\cite{science.oringin_2000,ahn2018dirac,zhoushuyun2018,PhysRevB.97.165414_2018,yang2018visualizing,li2023recent,lisi2021observation,RevModPhys.93.025006_2021,zhang2022angle,PhysRevLett.Xunan_2022,hamer2022moire,wang2023recent} and other techniques~\cite{li2024tuning}, even with a prominent departure from the commensurate approximation. This thus strongly implies that the concept of the energy band can be generalised into the incommensurate systems, without relying on Bloch's theorem.    

In this work, we theoretically illustrate the existence of a generalized concept of energy band for incommensurate systems in the absence of translational symmetry. 
 Such an important concept is rooted in the recently proposed incommensurate energy spectrum (IES) method~\cite{he2024energy}, enabling solving the eigenstates of an incommensurate system in momentum space with no need for Bloch's theorem. Here, we show that this approach provides a simple way to define the incommensurate energy band (IEB), which will recover the conventional energy band in a commensurate limit.  We further present an ARPES theory to calculate and interpret the ARPES spectra observed in incommensurate systems, which offers a direct way to detect the proposed IEB.  
As a demonstration, we first calculate the IEB, as well as the ARPES spectra, of the famous AAH model. Interestingly, it is shown that, with a proper period of the quasiperiodic potential, the AAH model is also able to host moir\'{e} flat bands, which may be the simplest moir\'{e} flat band system.  A 2D moir\'{e} system, i.e.~graphene under a triangle periodic potential with a $30^\circ$ twist angle, is studied as well, which exhibits complex IEB and ARPES spectra. 
This work presents a groundbreaking generalization of conventional energy band theory that transcends the confines of Bloch's theorem, thereby paving the way for many more in-depth understandings of the electronic properties of incommensurate systems. 


\begin{figure*}[t!]
    \centering
    \includegraphics[width=17cm]{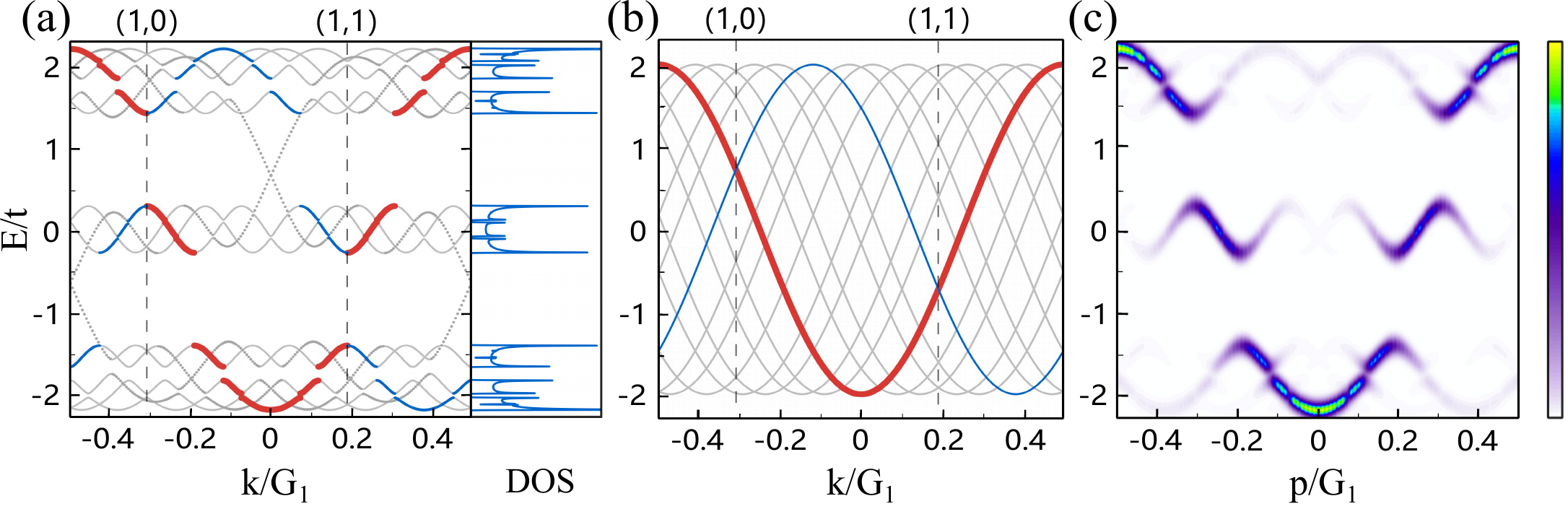}
    \caption{AAH model with $\alpha = (\sqrt{5}-1)/2$, $\vartheta = 0$. (a) is the calculated energy spectra and DOS with $V_0=1.2t$ and $n_c=6$. Red lines are the IEB denoted as $\epsilon^{(0)}(k)$. Blue lines highlight one of the replica bands. Dotted grey lines are the momentum edge states. Dashed lines denote the energy degeneracy points, represented as a pair of integers, where gaps open.  (b) is the energy spectra, similar to (a) but with $V_0=0$. The red line now is the pristine energy band of an atomic chain.  (c) is the calculated ARPES spectra $\sqrt{I}$ for (a), with a truncation $n_c=12$.} 
    \label{fig1}
\end{figure*}

\emph{Incommensurate energy bands}---To demonstrate the concept of IEB, we use the famous AAH model, a simple and intensively studied incommensurate model, as an example. The Hamiltonian is 
\begin{equation}
    H_{\mathrm{AAH}}=-t \sum_{j}( c^{\dagger}_{j}c_{j+1}+\mathrm{H.c.})+\sum_jV_jn_j 
\end{equation}
with  $V_j=V_0 \cos(G_2 \cdot r_j + \vartheta )$. Here,  $c_j$ is the annihilation operator at site j, $n_j=c^\dagger_{j} c_j$ is the number operator.  $V_0$ and $\vartheta$ are the amplitude and phase offset of the quasiperiodic potential, and $t$ is the nearest neighbour hopping. We set the lattice constant of the atomic chain to $a$ and the period of the quasiperiodic potential to $b$. Correspondingly, $G_2=2\pi/b$ ($G_1=2\pi/a$) is the magnitude of the reciprocal lattice vector of the quasiperiodic potential (atomic chain).  So, it becomes incommensurate once $\alpha \equiv G_2/G_1$ is an irrational number.

The energy band actually represents the eigenstates of the system classified by the momentum of electrons.  Thus, we first need to solve the eigenstates of the AAH model in momentum space, using the idea of  the recently proposed IES method. The great advantage of this method is that it does not require Bloch's theorem so that a truly incommensurate system can be treated in momentum space without resorting to the commensurate approximation. 

Since it is an atomic chain, the corresponding Bloch waves, $|k\rangle = \frac{1}{\sqrt{N}} \sum_j \exp(ik\cdot r_j) |j\rangle$, offer a more convenient basis, compared to the plane wave basis used in original IES theory. Then,  the Schrodinger equation in momentum space, $H(k) |\Phi (k) \rangle = E(k) |\Phi (k) \rangle$, becomes
\begin{equation}\label{Hk_maintext}
    \frac{V_0}{2} \phi_{m+1}(k) + \epsilon_{k+mG_2} \phi_m(k) + \frac{V_0}{2} \phi_{m-1}(k)=E(k) \phi_m(k),
\end{equation}
 where the Hamiltonian $H(k)$ is an infinite dimensional tridiagonal matrix~~\cite{supplemental1},
 $\epsilon_{k+mG_2} = -2t \cos[(k+mG_2)\cdot a]$ is the energy dispersion of the atomic chain. Note that  the eigenstates here are expressed as
 \begin{equation}\label{expression_wavefunction}
   |\Phi (k)\rangle = \sum_{m \in \mathbb{Z}} \phi_{m}(k) |k+mG_2\rangle
\end{equation}
where $\phi_{m}(k)$ is the coefficient of the Bloch basis. This is because, according to Eq.~\eqref{Hk_maintext},
each $|k\rangle$ is only coupled to the Bloch states in the set 
\begin{equation}\label{Qk}
    Q_k=\{ |k + m G_2\rangle: m \in \mathbb{Z}   \}
\end{equation}
where we always assume $k+mG_2$ represents $k+m G_2$ module $G_1$. In fact, as a Bloch wave,  $k+mG_2$ should be restricted in the FBZ of the atomic chain,  $[-\frac{G_1}{2}, \frac{G_1}{2}]$, named primary BZ (PBZ) in IES theory.  In other words, $H(k)$ is defined on the PBZ and a $k+mG_2$ should always be viewed as a point in PBZ.

In principle, provided a proper truncation of Hamiltonian matrix $H(k)$ is given, we can obtain the entire energy spectrum of AAH model by traversing $k$ across the whole PBZ and numerically diagonalizing $H(k)$. However, in this process, some momenta $k$ are calculated repeatedly.  This occurs because, $H(k)$ and $H(k+mG_2)$  exactly correspond to  the same matrix, thus they give rise to identical energy spectra.  It implies that, for the energy spectrum calculation, all the momenta in the set $Q_k$ are equivalent.  If we use the truncation $|m| \leq n_c$, there is a total of $N_E=2n_c+1$ equivalent momenta in the entire PBZ for a given $k$. Meanwhile, it can be proved that, for incommensurate systems, all the equivalent momenta in $Q_k$ will not coincide with each other~\cite{supplemental1}. 
It means that for a given momentum in PBZ, its energy spectra obtained from $H(k)$ have been repeatedly calculated $N_E$ times when $k$ traversing the whole PBZ. Therefore, any physical quantity of an incommensurate system can be calculated with~\cite{he2024energy}
\begin{equation}\label{IES}
\langle \hat{A} \rangle = \frac{1}{N_E}\sum_{k \in \mathrm{PBZ},i} p_{ki}  \langle \Phi_i(k)|\hat{A}|\Phi_i(k)\rangle
\end{equation}
which is the fundamental formula of the IES theory and the repeated calculating is solved by introducing a weight factor $1/N_E$. Here, $|\Phi_i(k)\rangle$ is the $i$th eigenstate of $H(k)$ and $p_{ki}$ is the Boltzmann factor. 
\begin{figure}[t!]
    \includegraphics[width=8.5 cm]{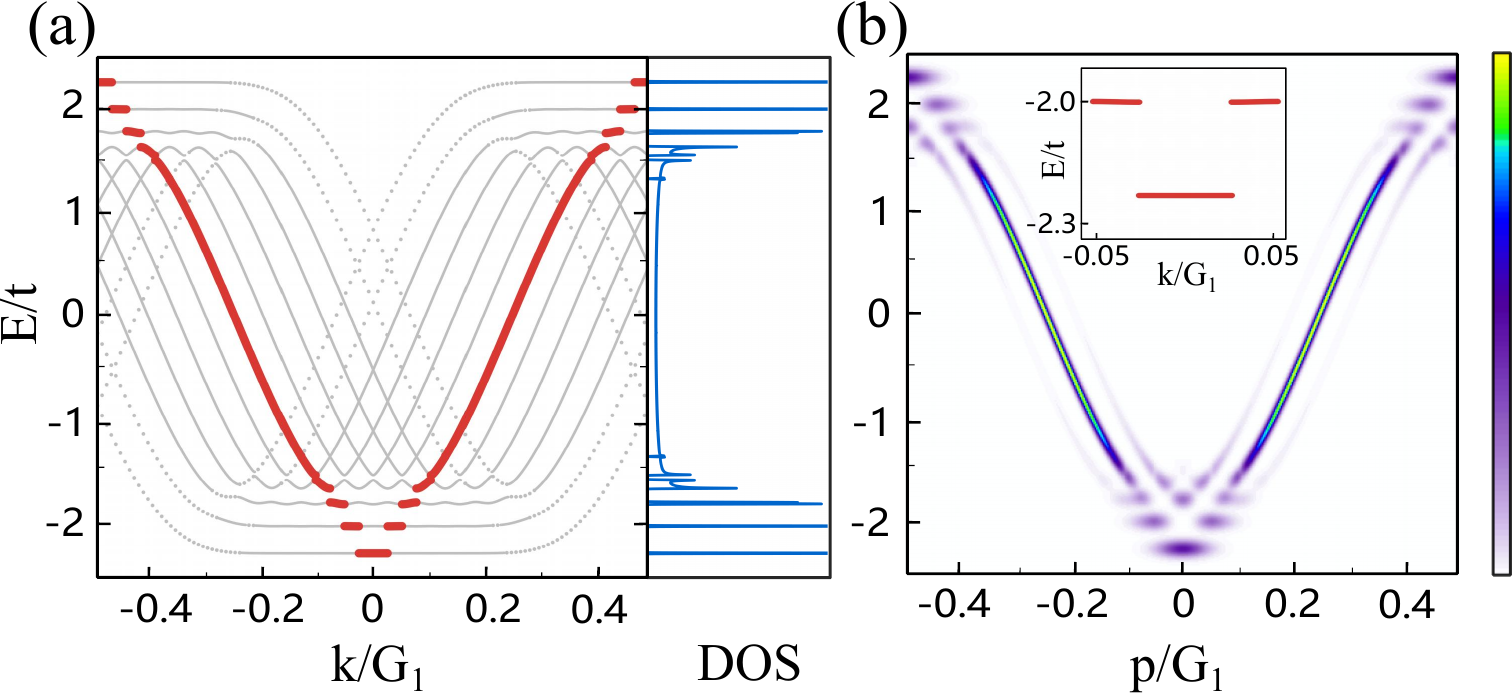}
    \caption{Moir\'{e} AAH model with $\alpha = \rm{sin}3^{\circ}$, $\vartheta = 0$. (a) is the calculated energy spectra and DOS with $V_0=0.4t$ and $n_c=6$. Red lines are the IEB denoted as $\epsilon^{(0)}(k)$.
     (b) is calculated ARPES spectra $\sqrt{I}$ for (a), with a truncation $n_c=12$. The inset is an enlarged view of the lowest two moire flat bands in (a). }
    \label{fig2}
\end{figure}

With the idea of IES theory above, the energy spectra of AAH model are plotted in Fig.~\ref{fig1} (a) as a function of $k$, with parameters: $\alpha=(\sqrt{5}-1) / 2$, $V_0 =1.2t$, $n_c=6$. And the corresponding DOS is given as well. Note that $m$ in Eq.~\eqref{Qk} in fact represents the order of the umklapp scattering from the quasiperiodic potential. For weak potential, a small truncation $n_c$ is able to give a sufficiently accurate result. For example,  $n_c=6$  here is a precise enough value. With these parameters, as illustrated above,   each eigenstate of the AAH model has been recalculated for $N_E=13$ times, resulting in a very complex energy spectra structure in Fig.~\ref{fig1} (a).  

A well-defined energy band should include all the eigenstates of $H(k)$ and include them only once. 
Therefore, if the desired IEB does exist,  a critical task is to further identify the IEB of the AAH model from the calculated complicated energy spectra. Here, our significant finding is that the entire incommensurate energy spectra can be sorted into $N_E$ distinct groups with an equal number of states, and except for a few groups that involve so-called momentum edge states, each of the other groups includes every eigenstate exactly once without any repetition~\cite{supplemental1}. 
We highlight two groups of the energy spectra, for example, see the red and blue solid lines in Fig.~\ref{fig1} (a). These groups of eigenstates have some typical features:
\begin{enumerate}
    \item  In each group, every momentum $k$ in PBZ uniquely corresponds to an eigenstate, allowing us to formally express its eigenenergy as a function of $k$, named dispersion of the eigenstates. 
    \item  In each group, the wavefunction of every eigenstate is predominately distributed on a specific  Bloch basis $|k+mG_2\rangle$. For example, the red group corresponds to $|k\rangle$ with $m=0$, while the blue one is associated with $m=1$. So, we can use the corresponding $m$  to label each group of eigenstates, whose dispersion is then denoted as $\varepsilon^{(m)}(k)$.
  \item A few groups of eigenstates with the largest $|m|$ contain some truncation dependent momentum edge states within spectral gaps, see the dotted lines in Fig.~\ref{fig1} (a), which are not the intrinsic eigenstates of the AAH model. Therefore, these groups should be excluded from calculations using Eq.~\eqref{IES},  leading to a revised $N_E$ as well~\cite{supplemental1}
    \item The remaining  $\varepsilon^{(m)}(k)$ have exactly the same shape, but they are shifted by different vectors $-mG_2$ on the PBZ. Considering the equivalence relations in $Q_k$, it implies that the recalculated eigenstates are precisely distributed into $N_E$ distinct groups, with each group containing exactly one of them.
    \item Near the energy degeneracy points of different Bloch basis, the corresponding dispersion curves will open up energy gaps, which divide the dispersion curves into several subbands. 
   
    
\end{enumerate}
These characteristics clearly show that the red group of eigenstates, denoted as $\varepsilon^{(0)}(k)$, fully meet all the requirements and expectations of the energy band:  like the conventional energy band, it includes every eigenstate of the AAH model exactly once, which can be expressed as a function of $k$ with a well-defined dispersion $\varepsilon^{(0)}(k)$. Therefore,  $\varepsilon^{(0)}(k)$ is actually just the IEB we are looking for, which can comprehensively describe all the electronic properties of the AAH model. 

\begin{figure*}[t!]
    \centering
    \includegraphics[width=16 cm]{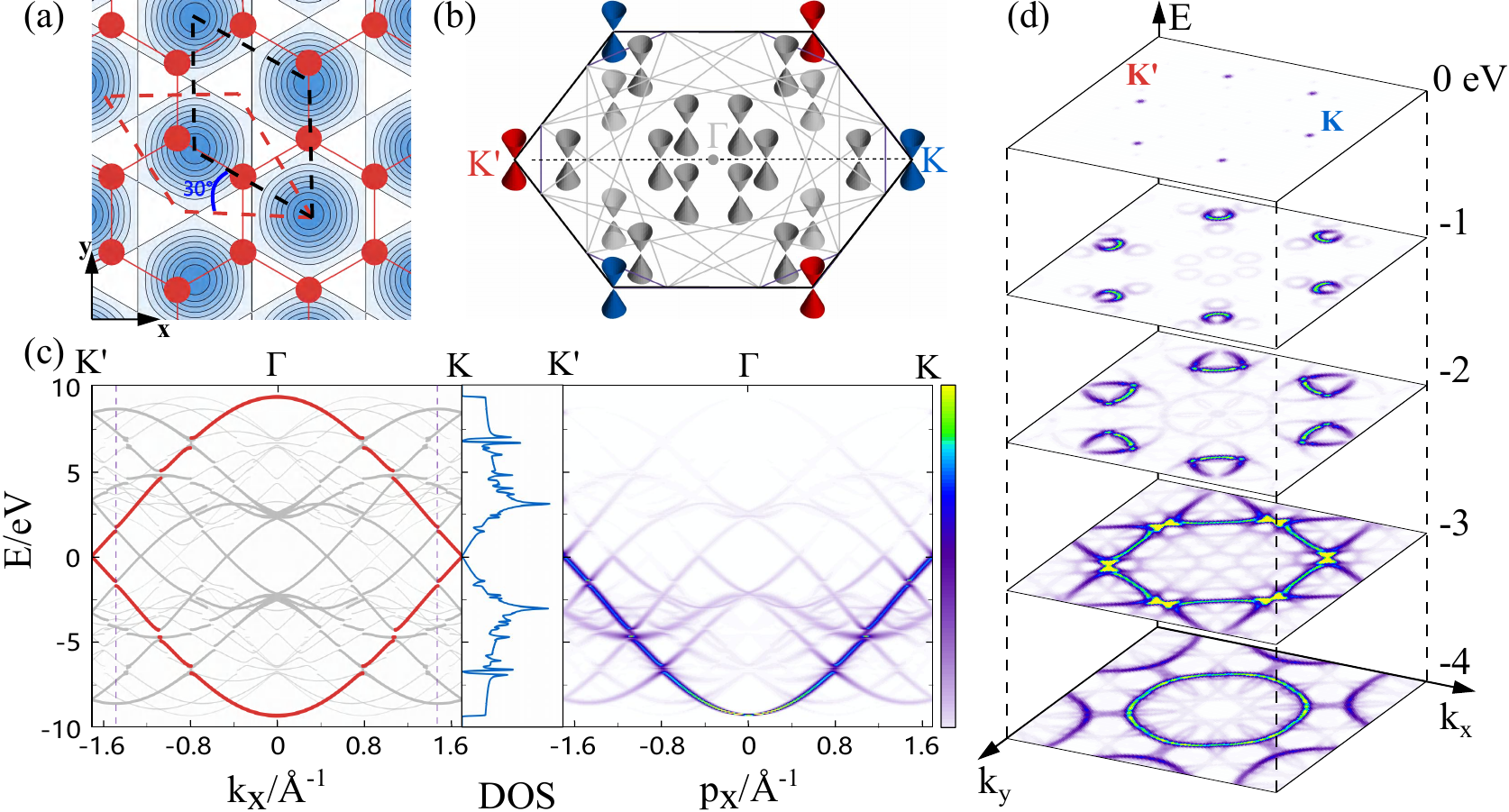}
    \caption{Graphene under a triangle periodic potential with a $30^\circ$ twist angle. (a) is the schematic of the potential. (b) illustrates the pristine bands of graphene (red and blue Dirac cones) and the replica bands (grey Dirac cones) in the case of $V_0=0$. (c) is calculated IEB (red lines), DOS and APRES spectra along the path $K'-\Gamma-K$ in $k$ space, with $V_0=268$ meV and a truncation $n_c=4$. (d) is the constant energy map of ARPES spectra with different binding energy from $0$ to $-4$ eV.}
    \label{fig3}
\end{figure*}

An intuitive understanding of the IEB can be gained by considering a weak quasiperiodic potential. When $V_0=0$, $H(k)$ becomes a diagonal matrix, where the eigenvalues are just the dispersion of each Bloch basis as plotted in Fig.~\ref{fig1} (b). The red line denotes the dispersion of Bloch basis $|k\rangle$, i.e.~$\varepsilon^{(0)}(k)$ with $m=0$, which becomes $\varepsilon^{(0)}(k)=\epsilon_k$ once $V_0=0$. And the blue line $\varepsilon^{(1)}(k)$ is similar, with  $\varepsilon^{(1)}(k)=\epsilon_{k+G_2}$ in the case of $V_0=0$. Similar as the conventional energy band theory, when a small $V_0$ is turned on, its influence can be treated by a perturbation method near the energy degeneracy point and a gap should occur. For instance, the energy degeneracy between the pristine and the other replica bands is determined by $\epsilon_k=\epsilon_{k+mG_2}$, which gives $k=(nG_1-mG_2)/2$ with an integer $n$ to ensure $k\in \mathrm{PBZ}$. As a result, the red and blue bands have two degeneracy points, each of which can be represented by a pair of integer $(m,n)$, i.e.~$(1,1)$ and $(1,0)$, as illustrated in Fig.~\ref{fig1} (a,b). Ignoring the influence of other bands, the size of the gaps at the two points opened by the quasiperiodic potential is $V_0$, which is a good approximation to the numerical results. Therefore, the quasiperiodic potential will open some gaps and then the pristine energy band in Fig.~\ref{fig1} (b) becomes the incommensurate one in Fig.~\ref{fig1} (a). In this sense, the IEB here can be regarded as a natural generalization of the extended zone scheme of the conventional energy band,  while other bands, $\varepsilon^{(m)}(k)$ with $m\neq0$ in Fig.~\ref{fig1} (a), are just some replica bands.  Interestingly, if the quasiperiodic potential becomes commensurate, the proposed IEB will revert back to the conventional ones~\cite{supplemental1}.
It once again confirms that the IEB, in fact, is a natural generalization of the energy band theory to the incommensurate systems.

We emphasize that the IEB  here does not rely on the truncation,   as long as 
a sufficient large $n_c$ is chosen. A larger truncation $n_c$ will involve more Bloch basis $|k+mG_2\rangle$ with larger allowed $|m|$. However, the effective coupling between the $|k\rangle$ and $|k+mG_2\rangle$ decreases for a larger $m$, leading to tiny gaps in $\varepsilon^{(0)}(k)$ which can not be detected in the experiment. So, we can test the convergence of the DOS~\cite{supplemental1} to ensure that a sufficient large truncation $n_c$ is achieved.

\emph{Incommensurate ARPES theory}---The concept of IEB proposed above can be further confirmed by ARPES measurements. As is well known, ARPES technique is the most direct way to detect the energy band structure in a solid, and it has indeed observed dispersing band states in incommensurate systems~\cite{science.oringin_2000,ahn2018dirac,zhoushuyun2018}, despite an ARPES theory for incommensurate systems is still lacking to date. The primary challenge lies in efficiently obtaining all the eigenstates of the incommensurate systems. Fortunately, our theory has successfully addressed this issue. Thus, an ARPES theory for the incommensurate systems can be developed straightforwardly now.    

Here, we use the simplest approximation where a free-electron final state is assumed~\cite{MOSER201729,sobota2021angle}. In the non-interacting case, the ARPES intensity of an incommensurate system, such as AAH model,  can be expressed as
\begin{equation}\label{ARPES_Intensity_universal_inc}
\begin{aligned}
         I(\mathbf{p},E) 
        \propto &|\mathbf{A}\cdot \mathbf{p}|^2\sum_{\mathbf{k}\in \rm{PBZ}}\delta[E- \varepsilon^{(0)}(\mathbf{k})]\times|\braket{\mathbf{p}|\Phi^{(0)}(\mathbf{k})}|^2\\
        =& |\mathbf{A}\cdot \mathbf{p}|^2 |\psi(\mathbf{p})|^2\sum_{\mathbf{k}\in \rm{PBZ}} \delta[E-\varepsilon^{(0)}(\mathbf{k})]\\
        &\times |\sum_{m}\phi^{(0)}_{m}(\mathbf{k})\delta_{\mathbf{p}_{||},\mathbf{k}+m\mathbf{G}_2+n\mathbf{G}_1} |^2
\end{aligned}
\end{equation}
where  $\mathbf{p}$ is the photoelectron momentum, $|\Phi^{(0)}(\mathbf{k})\rangle$ and $\varepsilon^{(0)}(\mathbf{k})$  represent the calculated IEB, e.g.~red line in Fig.~\ref{fig1} (a), and $\phi^{(0)}_{m}(\mathbf{k})$ is the corresponding coefficient of Bloch basis $|\mathbf{k}+m\mathbf{G}_2\rangle$, as given in Eq.~\eqref{expression_wavefunction}.  $\psi(\mathbf{p})=\int d\mathbf{r} e^{-i\mathbf{p}\cdot \mathbf{r}}\psi(\mathbf{r})$ is Fourier transform of the atomic orbital $\psi(\mathbf{r})$ (or Wannier orbital) on each site.  $|\mathbf{A}\cdot \mathbf{p}|^2 $ is a factor that describes the matrix element effects, relying on the polarization vector $\mathbf{A}$ of the incoming photon. 
A more detailed derivation is given in Supplemental Material (SM)~\cite{supplemental1}.

The ARPES intensity is proportional to the transition probability from the IEB $\varepsilon^{(0)}(\mathbf{k})$ to a final free electron state of the photoelectron with momentum $\mathbf{p}$ and kinetic energy $E_{kin}$. Energy conversation requires $E_{kin}=h\nu + E - \varPhi_W$, where $E$ is the energy of the initial state, $h\nu$ is the photon energy and $\varPhi_W$ is the work function. As ilustrated in Eq.~\eqref{ARPES_Intensity_universal_inc}, once $E=\varepsilon^{(0)}(\mathbf{k})$ is met, $I(\mathbf{p}, E)$ is nonzero only when $\mathbf{p}=k+m\mathbf{G_2}+n\mathbf{G_1}$, with a proper integer $n$ to make $k\in \mathrm{PBZ}$. And at each point $\mathbf{p}=\mathbf{k}+m\mathbf{G}_2+n\mathbf{G}_1$, $I$ is proportional to square of the corresponding coefficient on the Bloch basis $|\mathbf{k}+m\mathbf{G}_2\rangle$, which thus gives the maximum value when $\mathbf{p}=\mathbf{k}$ ($m=0$).  Therefore, the maximum value of the ARPES intensity will describe the shape of the IEB $\varepsilon^{(0)}(\mathbf{k})$, while there will also be weak signals on other replica bands $\varepsilon^{(m)}(\mathbf{k})$ ($m \neq 0$). These features are shown clearly in Fig.~\ref{fig1} (c), where the ARPES intensity of the AAH model is plotted. Here, $\sqrt{I(\mathbf{p},E)}$ is plotted in order to enhance the weak signals on the replica bands.

\emph{Moir\'{e} flat bands in AAH model}---The IEB theory offers a powerful theoretical tool for truly incommensurate systems. In the following, we show two interesting examples. The first one is the moir\'{e} AAH model, where the period of the quasiperiodic potential is very close to the lattice constant of the atomic chain. It is a difficult task to solve such an incommensurate system using the former real space method since it requires a huge supercell. However, the IEB method can readily solve this problem.  The numerical results are plotted in Fig.~\ref{fig2}, where we set $\alpha=\sin 3^\circ$ and $V_0=0.4t$.  The entire energy spectrum and the IEB (red lines) of such a model are shown in Fig.~\ref{fig2} (a). Interestingly,  moir\'{e} quasiperiodic potential produces several flat bands at the top and bottom, where the bandwidth of the lowest moir\'{e} band is less than $1\times 10^{-4}t$ with a large gap about $0.26t$~~\cite{supplemental1}. An enlarged view of the moir\'{e} flat bands is given in the inset of Fig.~\ref{fig2} (b). It indicates that the well-known AAH model may be the simplest system hosting moir\'{e} flat bands.  The corresponding ARPES spectra are calculated in Fig.~\ref{fig2} (b), which clearly reflects the shape of IEB. 

\emph{Graphene under an incommensurate potential}---Then, we show a 2D example, i.e.~graphene under an incommensurate potential. For simplicity, we consider a fictitious  triangle periodic potential with the same lattice constant but a
$30^\circ$ twist angle, see Fig.~\ref{fig3}~(a). The Hamiltonian is $H=H_{\textrm{G}}+V(\mathbf{r,\theta})$. Here,  $H_{\textrm{G}}$ is the tight binding Hamiltonian of graphene.  $V(\mathbf{r},\theta) = V_0\sum_{i=a,b,c}{\rm cos}[R(\theta)\mathbf{G}_i \cdot \mathbf{r}]$, where $\mathbf{G}_{a,b}$ are the two reciprocal lattice vectors of graphene and $\mathbf{G}_{c}=\mathbf{G}_{b}-\mathbf{G}_{a}$.  ${\rm R}(\theta)$ is the rotation matrix with $\theta=\pi/6$. We use $\Tilde{\mathbf{G}}_i={\rm R}(\pi/6)\mathbf{G}_i$ to denote the reciprocal lattice vectors of the $V(\mathbf{r})$.  
The calculation is similar to the AAH model. Using the Bloch waves of the A/B sublattice of graphene as the basis,  we first get a Hamiltonian matrix $H(\mathbf{k})$ defined on the FBZ of the graphene, i.e.~the PBZ of the IES theory. Then, the whole energy spectra and IEB can be calculated with the same process. Note that the energy spectra here possess all the properties enumerated above, which do not depend on the specific model.

The numerical results are shown in Fig.~\ref{fig3}.  Similar to the one dimensional case, when $V_0=0$, $H(\mathbf{k})$ becomes a block diagonal matrix, and its energy spectrum is composed of the pristine energy bands of graphene and some replica bands, as shown in Fig.~\ref{fig3} (b). Here, the replica bands (grey Dirac cones) result from shifting the pristine graphene bands (red and blue Dirac cones) by integer multiples of  $\Tilde{\mathbf{G}}_i$. The resulting IEB  and DOS are plotted in Fig.~\ref{fig3} (c), and the corresponding ARPES spectra are given as well. We further plot the ARPES images at different energies in Fig.~\ref{fig3} (d), which shows the details of the contour extending from low to high binding energy. 
This example clearly indicates that the concept of IEB, as well as the incommensurate ARPES theory,  is valid for 2D cases as well, thereby providing a powerful tool to understand the 2D incommensurate systems.  More detailed results of this example are given in SM~\cite{supplemental1}. 

\emph{Summary}---We summarize the standard process of obtaining the IEBs of any incommensurate system as follows: (1) Using the Bloch basis of the original periodic lattice, we obtain the Hamiltonian matrix $H(k)$ defined on PBZ; (2) Let $k$ traverse PBZ, and then diagonalize $H(k)$ to get the complete but redundant energy spectra; (3) The IEBs can be identified by examining the wave function distribution of eigenstates, since the wave functions of IEBs always predominantly reside on the Bloch wave of the pristine energy bands of the original periodic lattice.
The obtained IEB shares entirely analogous properties with the conventional energy bands, which includes every eigenstate of the incommensurate system exactly once, and thus can comprehensively describe all the electronic properties of the incommensurate systems. The proposed concept of IEB, as well as the incommensurate ARPES theory, is the central result of this work.

 Finally, we caution that the  IES here is only valid for delocalized regions of the incommensurate systems, i.e.~ the quasiperiodic potential should not be too large. 

 This work was supported by the National Key Research and Development Program of China (No.~2022YFA1403501), and the National Natural Science Foundation of China (Grants No.~12141401, No.~12174130, No.~12204044, No.~12474169).




\bibliography{reference}

\clearpage
\onecolumngrid

\newcommand{\bk}{\bm{k}}
\newcommand{\bq}{\bm{q}}
\newcommand{\btk}{\widetilde{\bm{k}}}
\newcommand{\btq}{\widetilde{\bm{q}}}
\newcommand{\br}{\bm{r}}
\newcommand{\cop}{\hat{c}}
\newcommand{\dop}{\hat{d}}
\newcommand{\xmark}{\ding{55}}
\def\Red#1{\textcolor{red}{#1}}
\def\Blue#1{\textcolor{blue}{#1}}

\begin{center}
\textbf{\large Supplementary Materials for: Energy Bands of Incommensurate Systems
}
\end{center}

\setcounter{equation}{0}
\setcounter{figure}{0}
\setcounter{table}{0}
\setcounter{page}{1}
\makeatletter
\renewcommand{\theequation}{S\arabic{equation}}
\renewcommand{\thefigure}{S\arabic{figure}}
\renewcommand{\bibnumfmt}[1]{[S#1]}

\section{I. The Hamiltonian matrix of the AAH model in momentum space}
As shown in Eq.~(2) of the main text, we obtain the Schrodinger equation of the AAH model, using the Bloch basis $(\cdots,|k-G_2\rangle, |k\rangle,|k+G_2\rangle,\cdots)$. Here, $|k\rangle = \frac{1}{N} \sum_j \exp(ik\cdot r_j) |j\rangle$ is the eigenstate of the atomic chain, and the corresponding eigenenergy is $\epsilon_k = -2t \cos(k\cdot a)$. Thus, the Hamiltonian matrix becomes an infinite dimensional tridiagonal matrix  
\begin{equation}\label{Hk}
    H(k)=\begin{bmatrix}
    &  \ddots            &               &            &                   &\\
          &\dfrac{V_0}{2} & \epsilon_{k+G_2}    & \dfrac{V_0}{2}&             &\\
          &              & \dfrac{V_0}{2} & \epsilon_k   & \dfrac{V_0}{2}   &\\
          &              &               &\dfrac{V_0}{2} & \epsilon_{k-G_2}&  \dfrac{V_0}{2}\\
          &              &               &              &                 &\ddots
\end{bmatrix}
\end{equation}

\section{II. The equivalent momenta of the incommensurate systems}
As mentioned in the main text, the relation $H(k)=H(k+mG_2)$ indicates that all the momenta in $Q_{k}=\left\{ q | q=k+mG_2: m \in \mathbb{Z} \right\}$  are equivalent for the energy spectrum calculation. For incommensurate systems, it should be noted that all these equivalent momenta will never coincide with each other in the PBZ. To prove this issue, let us assume that $q_m$ and $q_{m'}$ are two equivalent momenta in the $Q_k$
\begin{equation}\label{Set}
 \begin{aligned}
    q_{m} &= (k +m G_2) + n_m G_1,    \\
    q_{m'} &= (k + m' G_2)+ n'_{m'} G_1  
  \end{aligned}
  \end{equation}
  where  $m \neq m'$, and  $n_m$ and $n'_{m'}$ are two integers to ensure that $q_m$ and $q_{m'}$ are in PBZ.
   If the two momenta coincide, we have 
  \begin{equation}
    k +  mG_2 + n_{m}G_1  = k  + m'G_2 + n'_{m'}G_1.
\end{equation}
Then, we get
\begin{equation}
    \frac{G_2}{G_1} = \frac{n'_{m'}-n_m}{m - m'} = \alpha.
\end{equation}
It means that $\alpha$ is a rational number, which is contradict to the incommensurate potential.

\section{III. The structure of the incommensurate energy spectra}
\begin{figure}[ht]
    \centering
    \includegraphics[width=17 cm]{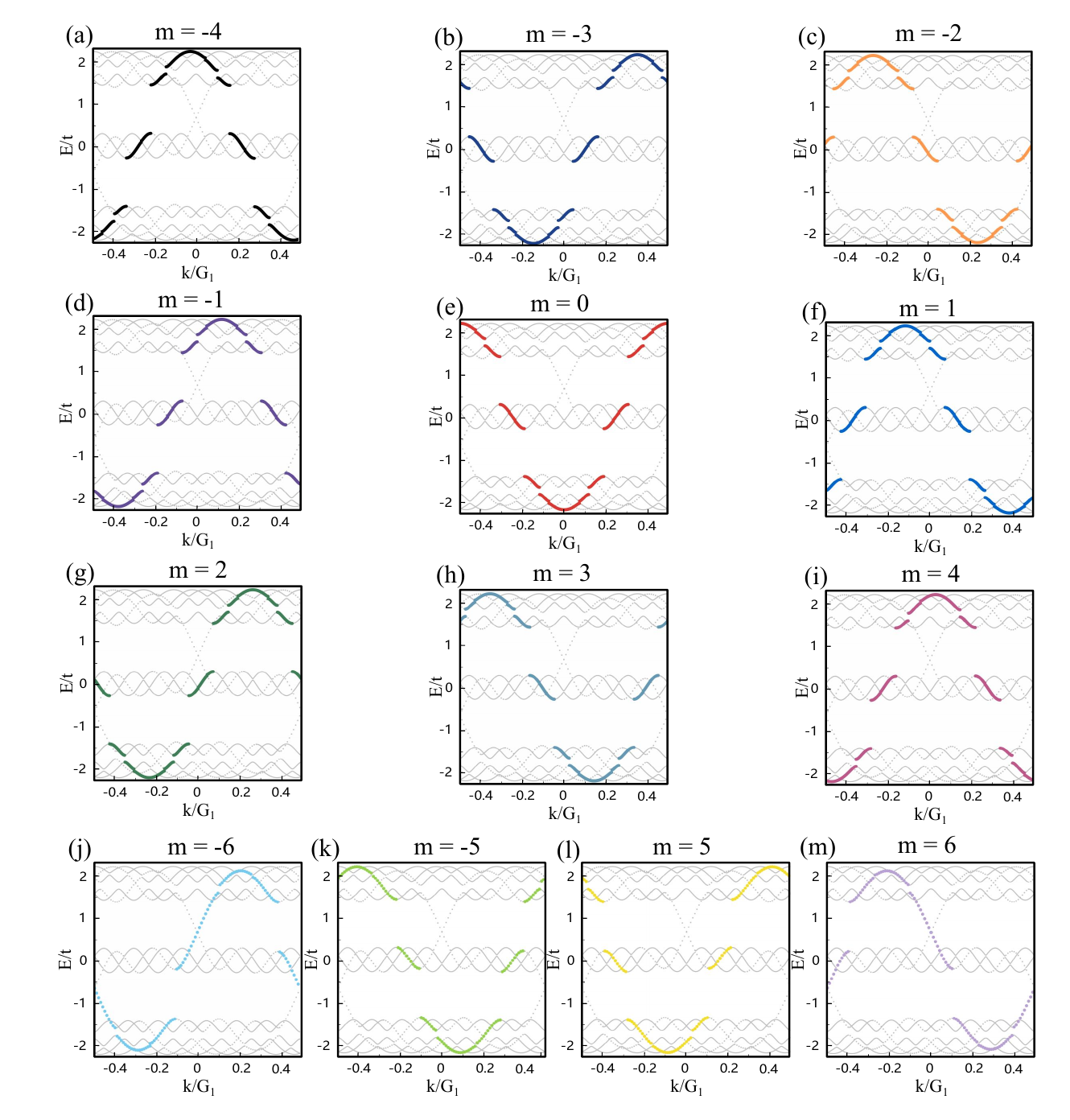}
    \caption{ AAH model with $\alpha = (\sqrt{5}-1)/2$. All the $\epsilon^{(m)}(k)$ are marked with different colors. (a)-(i): Red lines  in (e) represent $\epsilon^{(0)}(k)$, i.e.~IEB.  The other colored lines in  (a)-(d) and (f)-(i) correspond to the replica bands with the same shape. 
    (j)-(m): The colored lines highlight the replica bands involving momentum edge states. 
  Parameters: $n_c = 6$, $V_0=1.2t$, $\vartheta = 0$.
    }.
    \label{figs1}
\end{figure}

In the main text, we plot the incommensurate energy spectrum of AAH model in Fig.1, where the eigenstates are repeatedly calculated due the the equivalent relation. We find that the incommensurate energy spectra can be sorted into $N_E=2n_c+1$ groups with the equal number of states, as shown in Fig.~\ref{figs1}. Since the truncation $n_c=6$ is used here, we obtain a total of $N_E=13$ groups of eigenstates as plotted in Fig.~\ref{figs1}. Such classification is based on an important feature that, in each group, the wave function of the eigenstates is predominately distributed on a special Bloch basis $|k+mG_2\rangle$. Thus, we can use the value of $m$ (or the basis $|k+mG_2\rangle$) to label the groups of incommensurate energy spectra, see Fig.~\ref{figs1}. 

As shown in Fig.~\ref{figs1}, a dispersion $\varepsilon^{(m)}(k)$ can be formally defined for each group of eigenstates. In Fig.~\ref{figs1} (a)-(i), the dispersion $\varepsilon^{(m)}(k)$ have the same shape, but are shifted by different vectors $-mG_2$. Considering the equivalent relation, it indicates that all the eigenstates of the AAH model are included, and only included once in each of these groups. 

The groups with the largest $|m|$ involve so-called momentum edge states in the gaps, as shown in Fig.~\ref{figs1} (j)-(m). We give a detailed discussion in the next section. 

\begin{figure}[ht]
    \centering
    \includegraphics[width=14 cm]{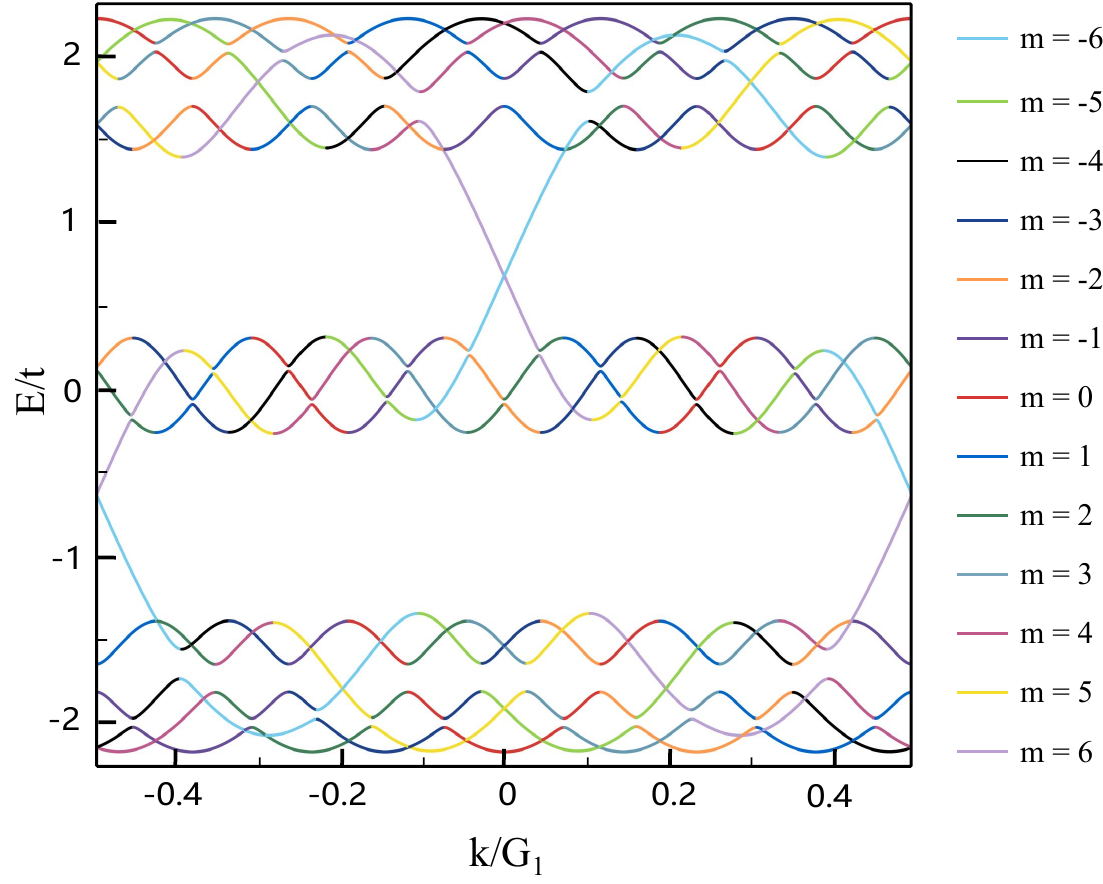}
    \caption{The structure of the incommensurate energy spectrum. The bands  $\epsilon^{(m)}(k)$ with different $m$ are marked with colors. The parameters used are the same as Fig.~\ref{figs1}.}
    \label{figs2}
\end{figure}

Finally, the structure of the incommensurate energy spectrum is illustrated in Fig.~\ref{figs2}, where different groups of eigenstates are labelled by color. 
\section{IV. Momentum edge states}
In Fig.~\ref{figs1} (j)$\sim$ (m), the groups with $|m|=5,6$ include so called momentum edge states in gaps. The momentum edge states result from the truncation of the Hamiltonian matrix. As shown in Eq.~\eqref{Hk}, the Hamiltonian is an infinite dimensional matrix. In order to calculate the energy spectrum, a truncation is needed. Formally, the Hamiltonian matrix can be viewed as a tight binding model in momentum space, where each site corresponds to a Bloch basis.  Therefore, a truncation will produce some edge sites in momentum space, which then give rise to some edge states.

Since these momentum edge states depend on the truncation, they are not the intrinsic eigenstates of the AAH model, which should be excluded from the calculation.  In Fig.~\ref{figs1}, we see that the groups with $|m|=5,6$ involve the momentum edge states. Once these groups are excluded, the revised $N_E$ should be $N_E=13-4=9$ in the calculation. 

\section{V. The commensurate limit}
\begin{figure}[ht]
    \centering
    \includegraphics[width=16cm]{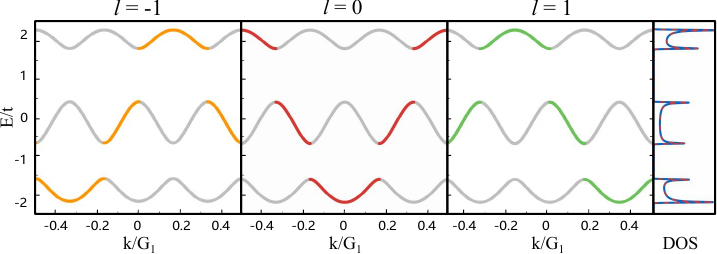}
    \caption{AAH model in the commensurate case with $\alpha=1/3$, $V_0=1.2t$, $\vartheta = 0$. (a)-(c) are  the energy spectrum, where $\epsilon^{(m)}(k)$ are marked with different colors. (d) is the corresponding DOS.
    }\label{figs3}
\end{figure}

Once the two periodic potentials become commensurate, the proposed IEB will revert back to the conventional energy bands. Here, we give an example of the AAH model with $\alpha=G_2/G_1=1/3$.  First of all, it should noted that the Hamiltonian matrix will become finite in the commensurate cases. Therefore, there is no need to perform truncation any more. The energy spectra are plotted in Fig.~\ref{figs3}, with different colors labelling the bands selected by the distribution of wave function. Now, the band $\varepsilon^{(0)}(k)$ (red lines) is just the extended zone scheme of the conventional energy band. 

\section{VI. The calculation details of the AAH model}
Here, we give some calculation details of the AAH model.

\begin{figure}[b]
    \centering
    \includegraphics[width=16 cm]{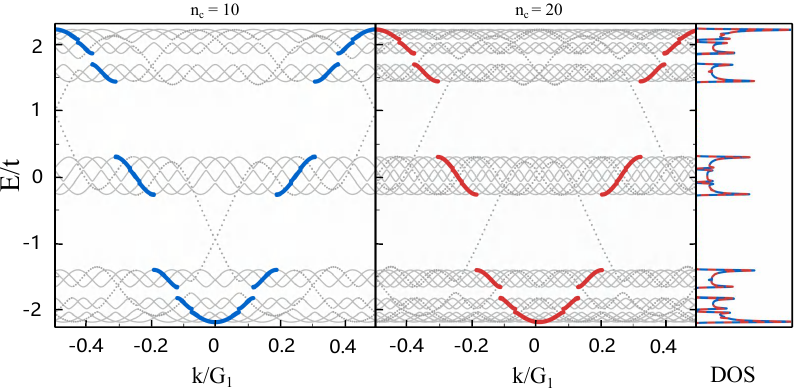}
    \caption{AAH model with $\alpha=(\sqrt{5}-1)/2$, $V_0 = 1.2t$, $\vartheta = 0$. (a)-(b) are the energy spectrum calculated with different truncation $n_c$. (c) are the corresponding  DOS.
    }.
    \label{figs4}
\end{figure}

\begin{figure}[h]
    \centering
    \includegraphics[width=12 cm]{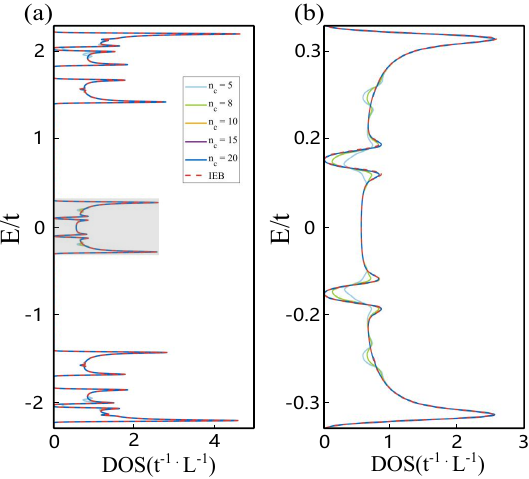}
    \caption{AAH model with $\alpha=(\sqrt{5}-1)/2$, $V_0 = 1.2t$, $\vartheta = 0$. (a) shows DOS calculated with different truncations, and the red dashed lines represent the DOS of the IEB. (b) are enlarged views of the grey regions in (a). 
    }.
    \label{figs5}
\end{figure}

Firstly, we show that the calculated incommensurate energy bands do not rely on the truncation, as long as the truncation is large enough. To illustrate this issue, we also test the convergence of truncation by comparing the DOS. The numerical results are plotted in Fig.~\ref{figs4}. Here, blue (red) lines represent the IEB calculated with truncation $n_c=10$ ($n_c=20$), and the corresponding DOS is compared as well with
\begin{equation}\label{DOS}
    \begin{aligned}
         {\rm DOS(E)} &= \frac{1}{N_E}\sum_{k \in {\rm PBZ}}\delta(E-E(k))\\
         & = \sum_{k \in {\rm PBZ}}\delta(E-\varepsilon^{(0)}(k)).
    \end{aligned}
\end{equation}
Clearly,  the two calculated IEB show no discernible difference, and their DOS are completely identical.   The comparison of DOS with different truncations are further shown in Fig.~\ref{figs5}, which indicates that $n_c=10$ is larger enough to get well converged results.

\begin{figure}[h]
    \centering
    \includegraphics[width=12 cm]{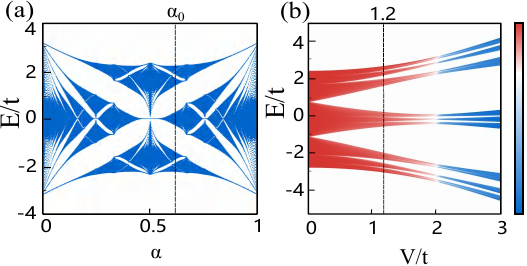}
    \caption{Hofstadter Butterfly spectrum and IPRM of AAH model. (a) is the Hofstadter Butterfly spectrum with $V_0=1.2t$, $n_c = 80$, $\vartheta = 0$. (b) shows the IPRM of AAH model with $\alpha=(\sqrt{5}-1)/2$, where ${\rm log_{10} (IPRM)}$ is represented by the color.
    }
    \label{figs6}
\end{figure}

Then, we illustrate that the calculated IEBs can correctly describe all the features of the AAH model. In Fig.~\ref{figs6} (a), we show that, as a function of $\alpha$,  the calculated IEB can produce the celebrated Hofstader butterfly spectrum of the AAH model. Meanwhile, as shown in Fig.~\ref{figs6} (b), the localization of AAH model can also be well described by calculating the inverse participation ratio in momentum space (IPRM) with IEB method~\cite{he2024energy,chen2021electrically}
\begin{equation}
    {\rm IPRM}(\ket{\Phi^{(0)}(k)}) = \sum_{m}|\phi^{(0)}_{m}(k)|^4 .
\end{equation}

\section{VII. The ARPES theory for incommensurate systems}
Using the proposed IEBs, an ARPES theory can be naturally formulated for the incommensurate systems, where the photoemission current I measured in an ARPES experiment can be expressed as 
\begin{equation}
    I(\mathbf{p},E) \propto  \sum_{\mathbf{k}\in {\rm PBZ}}|\braket{\mathbf{p}|H_{\rm{int}}|\Phi^{(0)}(\mathbf{k})}|^2 \delta(E- \varepsilon^{(0)}(\mathbf{k})).
\end{equation}
Here, we consider the non-interacting case, i.e.~sudden approximation,  and use the simplest approximation where a free electron final state $|\mathbf{p}\rangle$ is assumed~\cite{MOSER201729,sobota2021angle}.  $|\Phi^{(0)}(\mathbf{k})\rangle$ is the eigenstates of the IEB 
 \begin{equation}\label{wf}
 \begin{aligned}
   \ket{\Phi^{(0)} (\mathbf{k}) }= \sum_{m \in \mathbb{Z}} \phi^{(0)}_{m}(\mathbf{k}) \ket{\mathbf{k}+m\mathbf{G}_2},
 \end{aligned}
\end{equation}
where $\phi^{(0)}_m$ is the coefficient of the basis $|\mathbf{k}+m\mathbf{G_2}\rangle$. 

With the dipole approximation~\cite{sobota2021angle}, the Hamiltonian $H_{\rm{int}}$ describes the light-matter interaction:
\begin{equation}
    H_{\rm int} \sim \frac{e}{mc} \mathbf{A}\cdot\hat{\mathbf{p}}  = -\frac{ie\hbar }{mc} \mathbf{A}\cdot\nabla,  
\end{equation}
where $\hat{\mathbf{p}}$ is the photoelectron momentum operator, $\mathbf{A}$ is the vector potential of the incoming photon which is assumed to be constant in space.  Therefore, 
\begin{equation}\label{ARPES_IES}
\begin{aligned}
         I(\mathbf{p},E) \propto&  \sum_{\mathbf{k}\in {\rm PBZ}}|\braket{\mathbf{p}|H_{\rm{int}}|\Phi^{(0)}(\mathbf{k})}|^2 \delta(E- \varepsilon^{(0)}(\mathbf{k}))\\
        \propto &\sum_{\mathbf{k} \in {\rm PBZ}}|\braket{\mathbf{p}|\mathbf{A} \cdot \nabla  |\Phi^{(0)}(\mathbf{k})}|^2\delta(E- \varepsilon^{(0)}(\mathbf{k}))\\
        = &\sum_{\mathbf{k} \in {\rm PBZ}}|\mathbf{A}\cdot \braket{\mathbf{p}|\nabla  |\Phi^{(0)}(\mathbf{k})}|^2\delta(E- \varepsilon^{(0)}(\mathbf{k}))\\
        =&|\mathbf{A}\cdot \mathbf{\mathbf{p}}|^2 \sum_{\mathbf{k}\in {\rm PBZ}}|\braket{\mathbf{p}|\Phi^{(0)}(\mathbf{k})}|^2\delta(E- \varepsilon^{(0)}(\mathbf{k})),\\
\end{aligned}
\end{equation}
where we use $\nabla^{\dagger} = -\nabla$.
The inner product  can be directly calculated  
\begin{equation}
       \braket{\mathbf{p}|\Phi^{(0)}(\mathbf{k})} = \sum_{m}\phi^{(0)}_{m}(\mathbf{k})\braket{\mathbf{p}|\mathbf{k}+m\mathbf{G}_2}.
\end{equation}
The Bloch basis can be expended using the atomic orbitals,  $|k\rangle = \frac{1}{N} \sum_j \exp(i\mathbf{k}\cdot \mathbf{r}_j) |j\rangle$, where $|j\rangle$ is the atomic orbital of the $j$ site. We then obtain
\begin{equation}\label{eq_innerdc}
\begin{aligned}
    \braket{\mathbf{p}|\mathbf{k}+m\mathbf{G}_2} &=  \frac{1}{N}\sum_{j}e^{i(\mathbf{k}+m\mathbf{G}_2)\cdot \mathbf{r}_j}\braket{\mathbf{p}|j} \\ 
    &=\frac{1}{N}\sum_{j} e^{i(\mathbf{k}+m\mathbf{G}_2)\cdot \mathbf{r}_j} \int d\mathbf{r} e^{-i\mathbf{p}\cdot \mathbf{r}}\psi(\mathbf{r}-\mathbf{r}_j)\\
    &=\frac{1}{N}\sum_{j} e^{i(\mathbf{k}+m\mathbf{G}_2-\mathbf{p})\cdot \mathbf{r}_j} \int d\mathbf{r} e^{-i\mathbf{p}\cdot(\mathbf{r}-\mathbf{r}_j)}\psi(\mathbf{r}-\mathbf{r}_j)\\
     &=\frac{1}{N}\psi(\mathbf{p})\sum_{j}e^{-i\mathbf{p}\cdot(\mathbf{r}-\mathbf{r}_j)} \\
     &=\psi(\mathbf{p})\delta_{\mathbf{p}_{||},\mathbf{k}+m\mathbf{G}_2+n\mathbf{G}_1},
\end{aligned}
\end{equation}
where $\psi(\mathbf{p})=\int d\mathbf{r} e^{-i\mathbf{p}\cdot \mathbf{r}}\psi(\mathbf{r})$ and $\psi(\mathbf{r}-\mathbf{r}_j)$ is the wave function of the atomic orbital centered at $\mathbf{r}_j$. Finally, we obtain the Eq.~(6) in the main text: 
\begin{equation}\label{ARPES_final}
\begin{aligned}
         I(\mathbf{p},E) \propto|\mathbf{A}\cdot \mathbf{\mathbf{p}}|^2 |\psi(\mathbf{p})|^2\sum_{\mathbf{k}\in {\rm PBZ}}|\sum_{m} \phi^{(0)}_{m}(\mathbf{k})\delta_{\mathbf{p}_{||},\mathbf{k}+m\mathbf{G}_2+n\mathbf{G}_1}|^2\delta(E- \varepsilon^{(0)}(\mathbf{k})).
\end{aligned}
\end{equation}

\section{VIII.The calculation details of the moir\'{e} AAH model}
Here, we give some calculation details of moir\'{e} AAH model. 
\begin{figure}[h]
    \centering
    \includegraphics[width=16 cm]{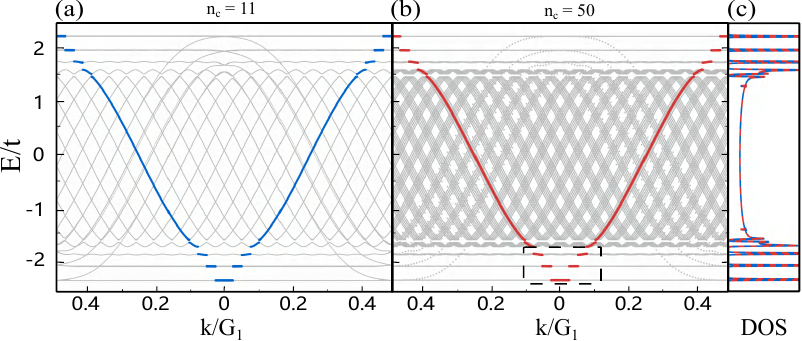}
    \caption{Moir\'{e} AAH model with $\alpha={\rm sin3}^{\circ}$, $V_0 = 0.4t$, $\vartheta = 0$.  (a)-(b) are the energy spectra calculated with different truncations, where the blue and red lines are the obtained IEBs.  (c) is the corresponding  DOS.
    }.
    \label{figs7}
\end{figure}
Firstly, in Fig.~\ref{figs7}, we plot the IEBs calculated with different truncation $n_c$. Then, in the enlarged view, we replot the moire flat bands near the band bottom, see Fig.~\ref{figs8}. We see that the bandwidth of the lowest moiré flat band is less than $10^{-4}t$ with a large gap of about $0.26t$.

Moreover, the numerical convergence with DOS is also illustrated in Fig.~\ref{figs9}. And Fig.~\ref{figs10} shows the corresponding IPRM results of the moir\'{e} AAH model.

\begin{figure}[h]
    \centering
    \includegraphics[width=14 cm]{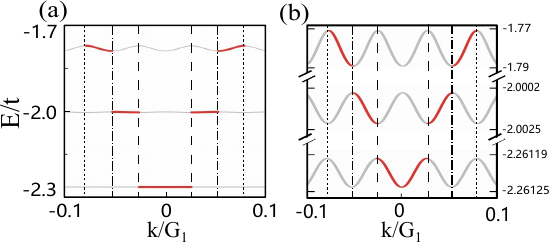}
    \caption{ The enlarged views of the Moir\'{e} flat bands in Fig.~\ref{figs7}.  The black dashed lines represent the Bragg planes.
    }\label{figs8}
\end{figure}


\begin{figure}[h]
    \centering
    \includegraphics[width=12 cm]{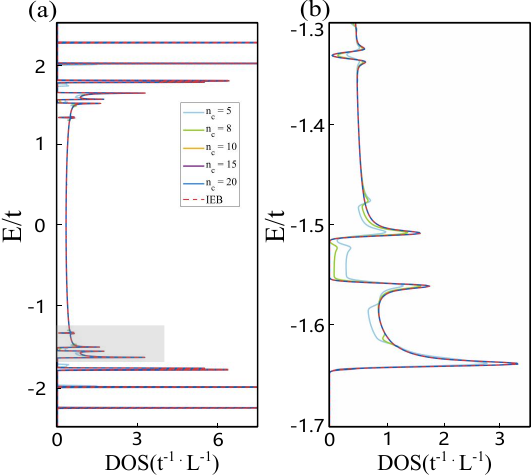}
    \caption{Moir\'{e} AAH model with $\alpha={\rm sin3}^{\circ}$, $V_0 = 0.4t$, $\vartheta = 0$. (a) shows the DOS calculated with different truncation $n_c$, and the red dashed line is the DOS of IEB. (b) are enlarged views of the grey region in Fig.~\ref{figs9}(a).}
    \label{figs9}
\end{figure}

\begin{figure}[h]
    \centering
    \includegraphics[width=8 cm]{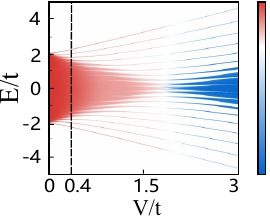}
    \caption{The IPRM of moir\'{e} AAH model with $\alpha={\rm sin3}^{\circ}$, $V_0 = 0.4t$, $n_c = 80$, $\vartheta = 0$. Colors represent ${\rm log_{10}(IPRM)}$.
    }
    \label{figs10}
\end{figure}

\section{IX. The calculation details of the graphene incommensurate model}
Here, we give some calculation details of the graphene incommensurate model, i.e.~graphene under an incommensurate potential.

The Hamiltonian of this model $H$  can be written as:
\begin{equation}
    H = H_{G} + V(r,\theta)
\end{equation}
where $H_G$ is the tight binding Hamiltonian of graphene.   $V(\mathbf{r},\theta) = V_0\sum_{i=a,b,c}{\rm cos}[R(\theta)\mathbf{G}_i \cdot \mathbf{r}]$ is the incommensurate potential.  $\mathbf{G}_{a,b}$ are two reciprocal lattice vectors of graphene, $\mathbf{G}_{a} = \frac{4\pi}{\sqrt{3}a_G}(0,1)$, $\mathbf{G}_{b} = \frac{4\pi}{\sqrt{3}a_G}(\frac{\sqrt{3}}{2},\frac{1}{2})$ with $\mathbf{G}_{c}=\mathbf{G}_{b}-\mathbf{G}_{a}$, $a_G = 2.46 \AA$ is the lattice constant of graphene and ${\rm R}(\theta)$ is the rotation matrix. In the main text, we define $\Tilde{\mathbf{G}}_i={\rm R}(\pi/6)\mathbf{G}_i$ as the reciprocal lattice vectors of the incommensurate potential. 
\begin{figure}[h]
    \centering
    \includegraphics[width=16 cm]{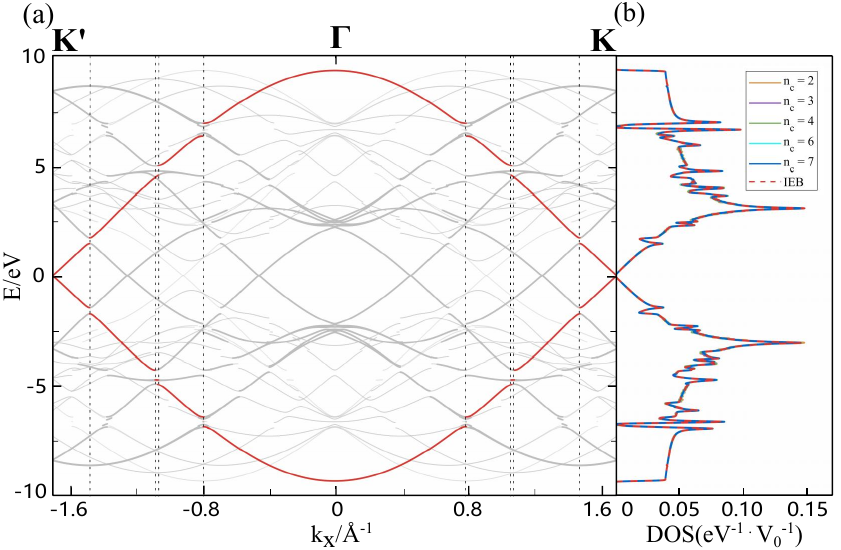}
    \caption{The graphene under an incommensurate potential with $\theta=\pi/6$, $V_0 = 268 {\rm meV}$. (a) is the energy spectrum of the graphene under an incommensurate potential, where red lines represent the IEBs. (b) is the DOS calculated with different truncation $n_c$, and the red dashed line is the DOS of  IEB.
    }\label{figs11}
\end{figure}
Similar to the AAH model, we then construct the Hamiltonian matrix in momentum space with $\ket{\mathbf{k}j}$ as the basis, where  $\ket{\mathbf{k}j}$ is the Bloch function of the graphene and  $j=A,B$ is the index of sublattice. The incommensurate potential only couples the Bloch waves in the set $Q_{\mathbf{k}}=\{ |\mathbf{k} + m_a \Tilde{\mathbf{G}}_a+m_b\Tilde{\mathbf{G}}_b,j=A,B \rangle:  m_{a,b} \in \mathbb{Z} \}$. We then use the truncation $|m_{a,b}| \leq n_c$. For a chosen value of $\mathbf{k}$, we can arrange all the coupled momentum in a certain order, so that the Hamiltonian matrix becomes a blocked matrix,

\begin{equation}
[H(\mathbf{k})]_{\mathbf{k},\mathbf{k}'} \equiv \left[
\begin{array}{cc}
    \braket{\mathbf{k},A|H|\mathbf{k}',A} &\braket{\mathbf{k},A|H|\mathbf{k}',B}  \\
   \braket{\mathbf{k},B|H|\mathbf{k}',A}  & \braket{\mathbf{k},B|H|\mathbf{k}',B}
\end{array}
\right]=\left\{
\begin{array}{rcl}
&h_{\mathbf{k}}  &{\rm if} \ \ \mathbf{k}'=\mathbf{k}.\\
    &\Delta &{\rm if}  \ \ \mathbf{k}'=\mathbf{k} -\Tilde{\mathbf{G}}_{a,b,c}. \\
    &\Delta^{\dagger} &{\rm if} \ \ \mathbf{k}'=\mathbf{k} +\Tilde{\mathbf{G}}_{a,b,c}. \\
    &0  &{\rm else}. 
\end{array}
\right.
\end{equation}

Here, $h_{\mathbf{k}}$ is the tight-binding Hamiltonian of graphene 
\begin{equation}
h_{\mathbf{k}} = 
\left[
\begin{array}{cc}
    0 &f(\mathbf{k})  \\
   f^*(\mathbf{k})  & 0
\end{array}
\right],
\end{equation}
where $f(\mathbf{k}) = -t\sum_{i=1,2,3}e^{i\mathbf{k}\cdot \mathbf{\delta}_{i}}$, and $\mathbf{\delta}_{1,2,3}$  are the relative position vectors between an atom and its three nearest neighbouring atoms in graphene.   $t$ is the NN hopping. And

\begin{equation}
\begin{aligned}
    \Delta =\frac{V_0}{2}
\left[
\begin{array}{cc}
    e^{i\mathbf{G}\cdot \mathbf{d}_A} &  0\\
    0& e^{i\mathbf{G}\cdot \mathbf{d}_B}
\end{array}
\right],\\
\end{aligned}
\end{equation}
where  $\mathbf{d}_A= \frac{a_G}{2\sqrt{3}}(0,-1)$ and $\mathbf{d}_B=\frac{a_G}{2\sqrt{3}}(0,1)$ represent the positions of the A/B atoms  in the unit cell.

The calculated IEBs for such graphene incommensurate model are replotted in Fig.~\ref{figs11}, where the red lines are the IEBs and other grey lines are replica bands.  The DOS calculated with different truncation $n_c$ is also given in Fig.~\ref{figs11}. We see that $n_c=4$ is sufficient to give a well converged DOS.

\begin{figure}[h]
    \centering
    \includegraphics[width=17 cm]{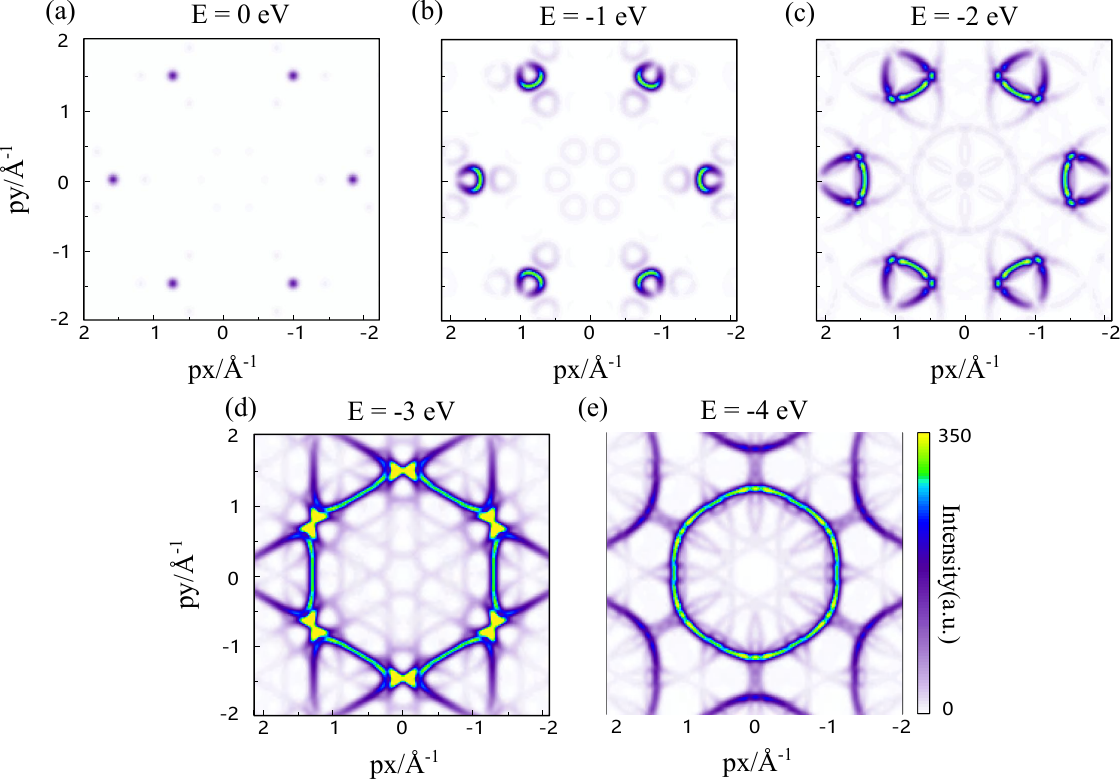}
    \caption{The constant energy map of ARPES spectra with different binding energy from $0$ to $-4$ eV. }\label{figs12}
\end{figure}

Then, we present the theoretical formula for the ARPES spectra in such graphene incommensurate system. 
The eigen wave function can  be expended using the Bloch basis
\begin{equation}
    \ket{\Phi^{(0)}(\mathbf{k})} = \sum_{\Tilde{\mathbf{G}},j=A,B} \phi^{(0)}_{\Tilde{\mathbf{G}}j}(\mathbf{k})\ket{\mathbf{k}+\Tilde{\mathbf{G}},j},
\end{equation}
where $\Tilde{\mathbf{G}}=m_a \Tilde{\mathbf{G}}_a+m_b\Tilde{\mathbf{G}}_b$, $m_{a,b}$ are integers with the truncation $|m_{a,b}| \leq n_c$. Thus, 
\begin{equation}
\begin{aligned}
    \braket{\mathbf{p}|\mathbf{k}+\Tilde{\mathbf{G}},j} &= \int d\mathbf{r}\braket{\mathbf{p}|\mathbf{r}}\braket{\mathbf{r}|\mathbf{k}+\Tilde{\mathbf{G}},j}\\
        &=\frac{1}{N}\sum_{\mathbf{R}} e^{i(\mathbf{k}+\mathbf{\Tilde{G}})(\mathbf{R}+\mathbf{d}_j)} \int d\mathbf{r}e^{-i\mathbf{p}\cdot\mathbf{r}} \psi(\mathbf{r}-\mathbf{R}-\mathbf{d}_j)\\
        &=\frac{1}{N}\sum_{\mathbf{R}} e^{i(\mathbf{k}+\mathbf{\Tilde{G}}-\mathbf{p})(\mathbf{R}+\mathbf{d}_j)}\psi(\mathbf{p})\\
        &=\psi(\mathbf{p}) e^{-i\mathbf{G}\cdot \mathbf{d}_j} \delta_{\mathbf{p}_{||},\mathbf{k}+\Tilde{\mathbf{G}}+\mathbf{G}}, 
\end{aligned}
\end{equation}
where $|\mathbf{p}\rangle$ is a plane wave. $ \psi(\mathbf{r}-\mathbf{R}-\mathbf{d}_j)$ is the atomic orbital of carbon atom centered at $\mathbf{R}+\mathbf{d}_j$, where $\mathbf{R}$ describes the position of unit cell.  $\psi(\mathbf{p}) = \int d\mathbf{r} e^{-i\mathbf{p}\cdot\mathbf{r}}\psi(\mathbf{r})$ is its Fourier transform.  Note that $\mathbf{G}$ is a reciprocal lattice vector of the graphene monolayer. Finally, taking photon polarization into account~\cite{PhysRevB.103.235146_2021}, the photoemission intensity can be calculated with 
\begin{equation}
\begin{aligned}
        I(\mathbf{p},E) 
        &=|\mathbf{A}\cdot\mathbf{p}|^2 |\psi(\mathbf{p})|^2\sum_{\mathbf{k}\in {\rm PBZ}}  |\sum_{\Tilde{\mathbf{G}},j} \phi^{(0)}_{\Tilde{\mathbf{G}}j}(\mathbf{k}) e^{-i\mathbf{G}\mathbf{d}_{j}}\delta_{\mathbf{p}_{||},\mathbf{k}+\Tilde{\mathbf{G}}+\mathbf{G}} |^2 \delta(E -  \varepsilon^{(0)}(\mathbf{k})).
\end{aligned}
\end{equation}
In Fig.~\ref{figs12}, we replot the ARPES data of the Fig.~3 (d) in the main text, in order to show more details.

\end{document}